\documentclass[]{elsart}
\usepackage{amssymb}
\usepackage{pst-node,pstricks,graphicx}
\begin{document}                
\flushright{}
\begin{frontmatter}
\title{Universal  lateral distribution of energy deposit in air showers
and its application to shower reconstruction}
\author[a]{D.~G\'ora,\corauthref{cor1}}
\author[b]{R. Engel,}
\author[b]{D. Heck,}
\author[a]{P.~Homola,}
\author[b]{H.~Klages,}
\author[a]{J.~P\c{e}kala,}
\author[a,b]{M.~Risse,}
\author[a]{B.~Wilczy\'nska,}
\author{and}
\author[a]{H.~Wilczy\'nski}
\corauth[cor1]{ {\it Correspondence to}: D.~G\'ora
(Dariusz.Gora@ifj.edu.pl)}
\address[a]{
Institute of Nuclear Physics PAN,
ul.Radzikowskiego 152, \\
31-342 Krak\'ow, Poland
}
\address[b]{
Forschungszentrum Karlsruhe, Institut f\"ur Kernphysik, 76021 Karlsruhe, Germany
}
\begin{abstract}
The light intensity distribution in a shower image  and its implications
   to the primary energy reconstructed by the fluorescence
technique are  studied.
 Based on detailed
CORSIKA energy deposit simulations, a universal  analytical formula  is derived for 
the lateral distribution of light in the shower image and a correction factor is obtained
to account for the fraction of shower light falling into outlying pixels in the detector.
The expected light profiles and  the corresponding correction of the 
 primary  shower energy  are illustrated 
for several typical  event geometries.
This correction of the shower energy can exceed 10\%, depending on shower geometry.
\end{abstract}
\end{frontmatter}
\section{Introduction}
\label{intro}
One of the methods of extensive air shower (EAS) detection is recording fluorescence light emitted 
by nitrogen molecules  in the  air along the shower path. 
For very high energies of the primary particle, enough fluorescence light
is produced so that the shower can be recorded from a  distance of many 
kilometers  by an appropriate optical detector system \cite{Bal,sommers1,sommers}.
As the amount of fluorescence light is closely correlated to 
the ionization energy deposit in air, 
it provides a calorimetric measure of the primary energy.

The field  of view of  a fluorescence detector (FD) telescope is divided into many  pixels.
For example,  in case of the  Pierre Auger Observatory (PAO) \cite{auger}
each pixel views $1.5^{\circ}$ of the sky and records the received  light 
in   100 ns time intervals.
A shower  passing through the  telescope field of view  triggers some  pixels, 
which form  together a  "shower track". 
The lateral width of this track depends on shower geometry but can well be larger than the
pixel size.    
 
For a precise energy determination one needs to collect the 
available signal as completely as possible, i.e. 
from all detector pixels which receive light from the shower.
On the other hand,  adding signals from many pixels implies adding the background noise as well.
Therefore it is important to include in the analysis only a small number
 of pixels which contain the ''true'' shower signal.




In this paper,   Monte Carlo simulations of the shower image  are presented.
 Based on  the spatial energy deposit of  shower particles as calculated by
CORSIKA~\cite{heck,markus3} it is shown that the lateral shower spread can  be well 
parameterized as a function of the shower age parameter only. The derived parameterization
can be used for reconstruction of shower profiles. 
This is illustrated by applying the new parametrization to the
  reconstruction of
the primary shower energy  for several  simulated  events in a  fluorescence detector.

The plan of the paper is the following: the definition of the shower width
 and algorithm of fluorescence light production based on the CORSIKA simulation of  
 energy deposit density  are described in Section 2. In Section 3 an analytical
 parametrization is derived  and  its implementation in energy reconstruction procedure 
is discussed. Conclusions are given   in Section 4.

\section{Properties of shower image}
\subsection{Shower width and shape function}

Photons which constitute an instantaneous image of the shower
originate from a range of shower development stages \cite{dgora}, namely  from the  surface $S$  shown in Figure 1. 
These simultaneous photons are defined as those which arrive at
 the FD during                                                                  
a short time window $\Delta t$.
During this $\Delta t$ (corresponding to a small change of the shower position in the sky by                    
$\Delta \chi$)  the shower front moves downward along the shower         
axis by a small distance $\Delta l \simeq R\Delta\chi/\sin(\theta_i)$, where $R$ is the distance from FD to the volume $\Delta V$
and $\theta_i$ is the angle between the shower axis and the direction towards FD.
This means that the small element of surface $S$                                                                          
corresponds  to a small volume $\Delta V$.                                                                  
The number of photons which   arrive  to  the FD from volume $\Delta V$                                                                        
can be calculated as:
\begin{equation}                                                                                                                               
\label{eq-fluo}                                                                                                                                
dN^{rec}_{\gamma}|_{\Delta V}= f(X,r) dS_{\perp}   \times                              
 \frac{ A }{4\pi R^{2}} \times \int W(\lambda) \eta(\lambda) d\lambda,                                                                                                              
\end{equation}  
 
where $f(X,r)$ is the  distribution of light emitted,                                                            
$dS_{\perp}$ is the  projection of the surface  $dS$ onto a surface
perpendicular to direction of the shower axis,
 $A$ is the light collecting  area of the detector, 
 $W(\lambda)$ is  the light 
 transmission factor, $\eta(\lambda)$  is the normalized fluorescence wavelength spectrum.
These photons form an instantaneous image of the shower which can be described by 
the angular distribution of light recorded by the FD

\begin{equation}                                                                                                                               
\label{eq-dis}                                                                                                                                
f_{\gamma}(\alpha)\equiv \frac{dN^{rec}_{\gamma}|_{\Delta V}} {\alpha
d\alpha d\phi },
\end{equation}  
 
where $\alpha$ is the small angle between the direction to the center of the image spot and
the  direction to volume $\Delta V$,
 $\phi$ is the  azimuth angle.   

The size of shower image is defined as the minimum angular diameter $2\alpha$ of the image
spot containing a certain fraction $F(\alpha)$ of the total light recorded by the FD.

A shower viewed from  a large distance has, to a very good approximation,
a circular image, independent of the direction of shower axis \cite{sommers1,dgora}.
The intensity distribution of light in this image is proportional to the lateral
distribution of the emitted fluorescence light in the shower at the viewed stage of 
evolution. Therefore  the  fraction of light received  $F(\alpha)$ can be obtained from
the corresponding  fraction $F(r)$ of light  emitted around the shower axis
\begin{equation}
\label{eq-shz}
 F(\alpha)\equiv\int_{0}^{\alpha} f_{\gamma}(\alpha^{'})2 \pi \alpha^{'} d\alpha^{'} \sim \int_{0}^{r} f(r^{'})2\pi r^{'} dr^{'} \equiv F(r),
\end{equation}
where $f(r)$ is the (normalized) lateral distribution  of fluorescence light emitted.
Here we have neglected the fact that photons from the side of the shower front
facing the detector have been emitted at a time later than photons
from the farther side of the shower. As will be shown later, relevant lateral
 distances are in the range 150 to 300 m and hence this is also the distance scale for
 the maximum time differences of emission \cite{sommers1}. As there is no significant change of the lateral distribution 
 expected of a shower traversing $\sim$300 m of air, the effect of different emission is negligible 
 in our calculation.  

In the following we shall consider showers close to the FD. For these
showers, the optical image size is  mainly determined by the geometric size
of the shower disk. Light absorption and multiple scattering  cause
only a minor, negligible modification of the shower image.

The main task is therefore to derive $f(r)$,
which is also referred to as the shape function,
since the  brightness distribution of the shower image
depends on the shape of $f(r)$.
At the first approximation, the $f(r)$  is proportional 
to the number of particles  in the shower at a given
lateral distance, assuming a constant fluorescence yield per particle in the shower.
  For  an electromagnetic shower
 this number of charged particles   is
  given by  the Nishimura-Kamata-Greisen (NKG) function \cite{nkg}. In this
 case  the  $F(r)$ can  be determined analytically, 
as has been shown in  Ref. \cite{gora-last}. However, 
$F(r)$ derived in this way
does not describe the light distribution well in case of hadronic showers. This is due
to the fact that the number of particles   in  a hadronic shower does not 
follow the NKG distribution well.
\subsection{CORSIKA approach based on energy deposit density}
\label{cor-prod}

Although the assumption that the  amount of emitted light is proportional
to the number of particles is adequate 
for a determination of the fluorescence 
signal in many cases, it is here not suited for the following reasons.                                                                                                                
The fluorescence yield is  proportional to the ionization energy deposited                                                                            
by the shower rather than  to the total number of charged particles. Furthermore,                                                                         
the simulated number of particles in a Monte Carlo calculation                                                                                   
depends on the threshold energies chosen by the user \cite{risse1,markus3}, 
above which particles are simulated. Particles falling below the threshold energy 
are discarded.

A better approximation for the fluorescence yield can be obtained by 
using   the   energy deposit  $dE(X)/dX$ as a function of  atmospheric slant depth
interval $dX$                                                                                        
together with a density- and temperature-dependent fluorescence yield                                                                            
$Y(\rho,T)$ \cite{kakimoto,nagano,nagano1}.                                                                                                                                    
In this approximation the distribution of photons emitted around the shower axis                                                                             
is proportional to the lateral  distribution of energy deposit, $f(r)\sim \frac{dE(X,r)}{dX_{v}}$, where 
$dX_v=dX\cos(\theta)$ is the vertical  depth interval and $\theta$ is the shower zenith angle.
The distribution of  energy deposit $dE(X)/dX_{v}$ is calculated with
the CORSIKA shower simulation program                                                                                                                                                                                                                 
as the sum of the energy released by charged                                                                                                     
particles  with energies above the simulation threshold plus the releasable energy                                                                              
fraction of particles discarded due to the energy cut. More specifically, the following
approximation is used~\cite{markus3}:
\begin{equation}
\frac{dE(X)}{dX_{v}}=\frac{E_{ioniz}}{\Delta X_{v}}+\frac{E_{e^{\pm}, cut}}{\Delta X_{v}}
+\frac{E_{\gamma, cut}}{\Delta X_{v}}+\frac{1}{3}\frac{E_{\mu^{\pm}, cut}}{\Delta X_{v}}
+\frac{1}{3}\frac{E_{had,cut}}{\Delta X_{v}}
\end{equation} 
where $E_{ioniz}$ is the ionization energy deposit of all charged particles traversing the depth interval
$\Delta X_{v}$. $E_{i, cut}$
 denotes the energy of particles of 
 type $i$ falling below the simulation threshold within this  interval.

In the following we will study the lateral distribution of energy deposit density 
in air showers, as it is directly proportional to the number of expected fluorescence photons.
Using CORSIKA a  two-dimensional energy deposit distribution around the  shower axis is
stored in histograms during the simulation process for  20 different vertical atmospheric depths.
Each of the 20 horizontal layers has a thickness of $\Delta X_{v}=1$ g/cm$^{2}$ and 
corresponds to a certain atmospheric depth:  the 
first one to $X_{1}=120$ g/cm$^{2}$  and the last one to  $X_{20}=870$ g/cm$^{2}$.
Linear interpolation between the observation levels 
is performed  in order to get the lateral distribution at a  given  vertical depth $X_n$   
located between two CORSIKA observation levels $X_{k}$ and $X_{k+1}$.
The  fraction of energy deposit $F(r)$ is calculated by numerically
integrating the histograms up to the lateral distance $r$.

The shower simulations are performed with the hadronic interaction models 
GHEISHA \cite{gheisha} (for interactions below 80 GeV) and QGSJET 01                                                                                
\cite{qgsjet}. Electromagnetic interactions                                                                                             
are  treated by a customized version of the  EGS4 \cite{egs4,egs4a}  code.                                                                             
To reduce computing time, a thinning algorithm \cite{hillas}                                                                                      
is selected within CORSIKA. The thinning level of $10^{-6}$ has been chosen with                                                                 
the so-called optimum weight limitation~\cite{kobal,risse}.                                                                                      
This ensures that the artificial fluctuations in the longitudinal shower                                                                          
profiles introduced                                                                                                                              
by the thinning method are sufficiently small for this analysis~\cite{risse1}.                                                                   
The kinetic energy thresholds for explicitly tracking particles                                                                                  
were set to: 100, 100, 0.25, 0.25 MeV for hadrons, muons,                                                                                      
electrons and photons, respectively.


\section{Results}

The knowledge of the  $F(r)$ function can be  used  to  calculate  
the ''true''  signal (light) from shower, which may be divided among several neighboring 
 detector pixels.
Below we propose a universal parameterization  of $F(r)$ based on CORSIKA simulations
and  show how the ''true'' signal can be  estimated with this parameterization.
\subsection{ Fraction of energy deposit density  from CORSIKA}


In the following we study the dependence of the lateral energy deposit   density on various variables.
 
 A natural transverse scale length in air showers, which proves to be useful for obtaining a universal
 parameterization of the  lateral distribution, is given by the Moli\`ere radius  \cite{molier}

\begin{equation}
\label{eq-mol1}
r_{M}\equiv E_{s}\frac{X_{l}}{\epsilon_{0}}, 
\end{equation}

where $E_{s}\simeq21$ MeV is the scale energy, $\epsilon_{0}=81$ MeV  the critical energy and $X_l=37$ g/cm$^2$ 
 the radiation length in air. The local Moli\`ere radius at a given atmospheric 
  depth of shower development 
 (at altitude $h$)  can be obtained by dividing Eq.~(\ref{eq-mol1}) by the air density, $\rho(h)$, and
is approximately given by $r_{M}= 9.6 \mbox{gcm$^{-2}$}/\rho(h)$.

It is also  well known that  the distribution of
particles  in a shower at a given depth depends  on the history of
the changes of $r_{M}$  along the shower path rather  than on the local $r_{M}$ value
at this depth. {To take this  into account,   the
$r_{M}$ value  is calculated  at $2$ cascade units (radiation length $X_{l}$) 
above the considered depth  \cite{molier}}. 
Using the   value of  the  Moli\`ere radius  calculated based on
the  atmospheric profile (the US Standard Atmosphere \cite{us})
for vertical depth  $X_{n}-2X_{l}\cos(\theta)$,  the  fraction of energy deposit density
$F(r^{*})$ versus the distance  in Moli\`ere units $r^{*}=r/r_{M}$ is found.
The knowledge of $F(r^{*})$ gives a possibility to study variation of the shape
of energy deposit density  due to properties of the atmosphere. 
The variation  of the density of the atmosphere along the path of a shower affects
the Moli\`ere radius and consequently also the radial particle distribution.

To characterize the development stage of a shower, we introduce 
  the  shower  age parameter 
\begin{equation}
s = \frac{3}{1+ 2X_{max}/X}, 
\end{equation}
  where $X_{max}$ is the atmospheric depth of shower maximum extracted from simulated
data\footnote{$X_{max}$  was determined by fitting a Gaisser-Hillas type function \cite{long} to the 
CORSIKA longitudinal profile of energy deposit.}.   With this definition, a shower reaches its
maximum at $s=1$.


Figure~\ref{fig339int} presents the integrals of energy deposit density $F(r)$ 
and $F(r^{*})$ 
for a vertical  proton shower with primary energy $E_{0}=10$ EeV, 
obtained  at  different  atmospheric depths.
It is seen in Figure~\ref{fig339int}A that the shape of this integral 
 distribution varies considerably only at depths
 smaller than  360 g/cm$^{2}$. At larger depths,  and in particular around the shower maximum,
 the variation of   integral   energy deposit  profile is  not significant. 
 However,  this  variation is  larger when one  plots this  integral  
  versus distance measured in Moli\`ere units, as shown in Figure~\ref{fig339int}B.

Figure~\ref{fig-en}A shows the dependence  of the integral of energy
 deposit density $F(r)$  on energy and primary particle. 
It is seen that the integral  profile only slightly depends on energy and primary particle. 
The differences are even smaller if  we plot the 
 fraction of energy deposit density  versus distance in Moli\`ere  units, 
as shown in Figure~\ref{fig-en}B. 
The same shape of the  $F(r^{*}$)  profile for different primaries and energies  means 
that variations of the  $F(r)$ profile  are mainly due to  the   atmospheric effect
i.e. dependence of the Moli\`ere radius on altitude.
For the same shower geometry, there are different altitudes of the maxima of  proton and iron  showers,
and in consequence different  values of Moli\`ere radius $r_{M}$. 
Since $r_{M}$ determines the lateral spread
of particles in the shower,  the shape function $F(r)$ becomes broader
for iron showers (higher altitude of shower maximum 
than for proton shower).

  Figure~\ref{fig-inc} presents the dependence of the integral of  
  energy deposit profile on  zenith angle.
   We note that CORSIKA energy deposit lateral profiles are obtained 
   for  horizontal planes  at the given  observation level, so  
   if one  compares energy deposit densities between vertical and inclined showers, 
   a projection of densities from horizontal plane
     to the plane normal to the shower axis is performed for inclined showers.  
   The corrected profile  for a shower inclined  at $\theta=45^{\circ}$ 
   is  shown in Figure~\ref{fig-inc}A by the solid line. It is seen that this profile
   differs from the profile obtained for a vertical shower.  
   This means that the shape of $F(r)$ depends  on the zenith angle.    
   This dependence can  be understood  if one  takes  into account the  influence of  
   the atmospheric effect on the energy deposit profile $F(r)$.
    For a homogeneous atmosphere, the  shape function for inclined and vertical  showers 
 must be  the  same for the same development stage because the  Moli\`ere radius does 
 not change  with
 altitude. In case of an inhomogeneous atmosphere,  differences of the shape function between vertical and
 inclined shower  should be  proportional to the differences of the Moli\`ere radius
  (i.e density of air). Thus if changes of the shape function 
   are due only to the  atmospheric effect,
 then  $F(r^{*})$  profile should be the same  for  vertical  and  inclined showers.
 Figure~\ref{fig-inc}B confirms this assumption.

The analysis of  Figs. 2B, 3B and 4B leads to the following conclusion: 
{\it  the lateral shape of the  energy deposit density 
versus distance from shower axis measured in Moli\`ere units is independent of the primary energy, 
primary particle type  and zenith
angle. It depends, to a good approximation,  only on the shower age}. 
Figure~\ref{fig9int}  confirms this conclusion, too.
In this  Figure we present the integral of the energy deposit density for different 
age parameters  for 10 individual proton 
and 5 individual iron showers with different zenith angles ($\theta=0^{\circ},45^{\circ}, 60^{\circ}$)
 and energy 10 EeV. It is seen that the shower-to-shower  fluctuations 
 are strongly reduced for a given age when we correct  $F(r)$ profiles for the 
   atmospheric effect i.e. plot $F(r^{*})$. Also, there are no  differences
 in the shape of $F(r^{*})$ for showers with different zenith angles and primary particle type.  
This means that it is  possible to  find a universal  function which   describes
 the shape of energy deposit density as a function of   shower age only. Following our   earlier
  work \cite{gora-last} we will use 
 the  function
\begin{equation}
\label{eq-fit}
F(r^{*})=1-\left(1+a(s)r^{*}\right)^{-b(s)},
\end{equation}
where the parameters $a(s)$ and  $b(s)$ are assumed to be   functions of shower age.
Fits of this  functional form  to the integral of energy deposit density  were performed for 
the data from  Figures~\ref{fig9int}B, D, F and  are shown in Figure~\ref{fit-age}.  
The values of the parameters  $a(s)$ and $b(s)$ for different  shower ages  are presented 
in Figure~\ref{fig-par}. The  age dependence of $a(s)$ and $b(s)$
parameters is well  described by   
\begin{equation}
\label{eq-fita}
a(s)=5.151s^{4}-28.925s^{3}+60.056s^{2}-56.718s+22.331,
\end{equation}

\begin{equation}
\label{eq-fitb}
b(s)=-1.039s^{2}+2.251s+0.676. 
\end{equation}

Thus, Eqs.~(\ref{eq-fit}), (\ref{eq-fita}) and (\ref{eq-fitb})  
 give us a universal function which describes 
the fraction of energy deposit density within a specified distance from the shower axis
 for different energies, zenith angles 
and primary particles. Moreover, Eq.~(\ref{eq-fit})
can be used to simulate the size of the shower image not only at shower maximum like 
in Ref.~\cite{gora-last}, but also for any shower development stage.  
Inverting  Eq.~(\ref{eq-fit}) 
and taking into account the distance from the detector to the shower ($R_{0}$)
we can find the angular size of the image $\alpha$ that corresponds to a certain
fraction of the total fluorescence light signal:
\begin{equation}
\label{eq-nkgshw}
\alpha(s)= 2 \arctan\left(\frac{r}{R_{0}}\right)=2
   \arctan\left(\frac{r_M(s)}{a(s)R_{0}}((1-F(r))^{-1/b(s)} -1)\right).
\end{equation}

\subsection{Application in energy reconstruction procedure}

As the shape of the lateral distribution of energy deposit  can be well described 
by  Eq.~(\ref{eq-fit}), 
it may be used  to take into account  the knowledge on the shower
width in  the procedure of shower reconstruction in the fluorescence detector.

One of the   first  steps in  shower energy  reconstruction  is the calculation  of the 
 light profile at the aperture of the 
detector, based on the signal recorded  by the detector pixels. 
This signal is converted to the number of     
 equivalent photons   at the detector diaphragm. For example, one procedure to determine such
  a profile is described in \cite{gapdawson}.
 This algorithm  uses  as input 
the reconstructed geometry to locate the  shower image  on the FD telescope  camera  in  100 ns intervals. 
Next,  the  signal (charge)  and noise from pixels lying within a predetermined  angular distance $\chi_{S/N}$ from the 
instantaneous position of the image spot center are collected to  find 
the radius  $\chi_{S/N}^{max}$ that maximizes the  signal-to-noise ratio {\bf over the whole} shower track.
 Finally, the charge in  each  100 ns time interval (time slot), $L_{S/N}(t)$, within that radius $\chi_{S/N}^{max}$ is found and converted 
 to the number of photons using calibration constants.

 This procedure  works well for distant showers,
when the light collected  within the 
radius $\chi^{max}_{S/N}$ corresponds to about 100\% of the true signal, 
but  some differences between the  signal within $\chi_{S/N}^{max}$ and the true signal may exist for nearby showers.
In the following we investigate this problem and  estimate
 a  correction  to  the described reconstruction algorithm. 

The necessary shower  reconstructions were performed using Flores-Eye~\cite{flores} 
and    FDSim~\cite{prado} programs. 
First, FDSim was used to generate    events   based 
on  the Gaisser-Hillas parameterization.
Next, the  geometrical and energy reconstruction were performed
using Flores-Eye. The reconstructed  geometries
  for some events are listed in Table \ref{tab1}.
 Using  these geometries,  we   find the collected signal $L_{S/N}(t)$
 within angular distance $\chi^{max}_{S/N}$ at each time interval\footnote{The value of $\chi^{max}_{S/N}$ depends on geometry, but
 for events listed in Table \ref{tab1} it equals about  $1.2^{\circ}$.}.  
Then, for the given $\chi_{S/N}^{max}$ the  effective radius around shower axis $r_{0}=R_{0}\tan(\chi_{S/N}^{max})$ 
and the fraction of light $F(r_{0})$  based on the function $F(r)$ was calculated.
The fraction $F(r_0)$  for events listed in Table \ref{tab1} is  shown in Figure~\ref{fig-frac}. 
It is seen that   for Event1  $F(r_{0})$ changes from   89\%  for a distance-to-shower
$R_{0}=7.0$ km to  87\%  for $R_{0}=6.0$ km.  
For other events,  the collected fraction of the signal within $\chi_{S/N}^{max}$   increases with  
increasing distance-to-shower and   equals on average  about 91\%, 94\% and 99\%  for  Event2, 
Event3, Event4, respectively. In  other words,  some portion of the  signal, 
which falls beyond $\chi_{S/N}^{max}$ is missing in the shower reconstruction procedure.
To take into account this lost portion of the signal,
the signal $L_{S/N}(t)$   is rescaled  according to formula $L_{true}(t)=L_{S/N}(t)/F(r_0)$.
In this way, one takes into account the  shape of  lateral distribution of energy deposit 
and obtains the  new integrated charge $L_{true}(t)$ for each time slot. Thus the part of the signal which was contained in 
neighboring pixels outside $\chi_{S/N}^{max}$ is accounted for.

It should be pointed out that in general, any reconstruction procedure has to  take into
account the pixellation of the detector. Independent of the specific  approach,
the correction procedure developed in this work can be applied.
  In the following, we demonstrate  the influence of this  correction on the light profile and 
  on energy determination. In Figure~\ref{fig-prof2}A it is shown that our correction  leads
   to considerable  differences 
    between the $L_{S/N}(t)$   (dashed line) and $L_{true}(t)$ profile (solid  line) 
    for a nearby shower (Event 1). 
In case of  distant showers (like Event4) the profile  is almost unchanged (see Figure~\ref{fig-prof2}B).
It should also be noted that changes in the  detector-to-shower distance are accounted 
 for in this approach. A ''differential'' correction is applied (i.e. for each time slot), which also leads to a better 
 reconstruction of the longitudinal shape (and thus $X_{max}$) of the shower.

Accepting only a fraction of the signal  contained within $\chi_{S/N}^{max}$ directly  influences the reconstructed  
 primary   energy of the shower. In Table \ref{tab2} we present the  influence  of the $L_{S/N}(t)$
  correction  on  the Gaisser-Hillas fit 
 to the reconstructed number of particles in the showers.
 It is  seen that this correction  changes both  the  number of particles at the  shower maximum
 and the position  of the  shower maximum. 
 These changes  lead to  different estimates of  primary energy.
In the last column  the relative differences $k_{E}=(E^{true}_{0}-E^{S/N}_{0})/E^{S/N}_{0} $ are listed. 
One sees that $k_{E}$ is always positive and decreases  from 14\% for a distance to shower maximum of
  $R_0$=6.5 km to 2\% for $R_0$=23 km.

\section{ Conclusions}

In this work,  the distribution of light 
in the shower optical image is analyzed, based on the lateral distribution 
of energy deposited by the shower,
as  derived from CORSIKA simulations.
 The lateral distribution of energy deposited is parameterized with
 a functional form inspired by  the NKG distribution.
 The angular distribution of photons arriving simultaneously at the detector 
 (i.e. the intensity distribution of light in the instantaneous image of the shower) is obtained.
  The shape of this distribution can be approximated by a universal function that  depends on the shower age only.

This universal function  is used to derive a correction to the shower energy due to the fraction of light falling 
into detector pixels located far from the center of the shower image. In the usual procedure of shower reconstruction,
signal-to-noise ratio is optimized, so that pixels lying far from the center of the shower
image are not included in the analysis. The percentage of shower signal in those outlying pixels was determined in this paper
based on the lateral distribution of light in the shower image.
The signal recorded by the 
fluorescence detector in the accepted central pixels is rescaled, so that a corrected light profile of the shower is obtained. 
 For events examined, this correction 
 increases the estimated shower energy by 2--14\%, depending on the detector-to-shower distance.

{\it Acknowledgements.}
This work was partially supported by the Polish
Committee for Scientific Research under grants  No. PBZ KBN 054/P03/2001  and 2P03B 11024
and in Germany by the DAAD under grant No. PPP 323.  MR is supported by the Alexander
 von Humboldt Foundation.

\newpage
\begin{table}[b]                                                                                                                  
\caption{Characteristics of events used for the comparisons in                                                            
this paper. The shower zenith angle $\theta$,                                                                                      
azimuth angle $\phi$, core position $X_c$, $Y_c$  are measured                                                                    
relative to FD detector.                                                                                          
\label{tab1}}                                                                                                                     
\vskip 0.5cm                                                                                                                      
\begin{center}

\begin{tabular}{ccccc}                                                                                                          
\hline                                                                                                                            
\hline                                                                                                                            
Event & $\theta$ &  $\phi$  & $X_c$ &  $Y_c$ \\                                                    
      & (deg)    &(deg)      & (km) & (km) \\                                                                    
\hline                                                                                                                            
 Event1 & 0. & -15& -0.820  & 5.860    \\
 Event2 & 60 & 175& 4.535 &6.906   \\
 Event3 & 45 & 20& 2.282 &7.066    \\
 Event4 & 45 & 50& -7.665 & 19.11    \\
                                                               
\hline                                                                                                                            
\hline                                                                                                                            
\end{tabular}
\end{center}                                                                                                                      
\end{table}   
                                                                                                                   

\begin{table}[]                                                                                                                  
\caption{Comparison of  Gaisser-Hillas function  parameters  based on the  $L_{true}(t)$
 and the   $L_{S/N}(t)$
light profiles and their influence on primary energy. 
The number of particles at shower maximum $N_{max}$,                                                                                      
corresponding slant depth  $X_{max}$, estimated energy $E_{0}$ and relative
difference $k_{E}=(E_{0}^{true}-E^{S/N}_{0})/E^{S/N}_{0}$  are listed as a function 
of   distance $R_0$  from detector  to shower maximum. Calculations are made using 
 the function  $F(r^{*})$  described by Eq.~(\ref{eq-fit}).                                                                                            
\label{tab2}}                                                                                                                     
\vskip 0.5cm                                                                                                                      
\begin{tabular}{ccccccccc}                                                                                                          
\hline                                                                                                                            
\hline                                                                                                                            
Event         & $R_0$    &  $N_{max}^{S/N}$ & $N_{max}^{true}$ &  $X_{max}^{S/N}$ &$X_{max}^{true}$&$ E_{0}^{S/N}$& $ E_{0}^{true}$& $k_E$ \\                                                    
              & (km) &  ($10^{9})$ & (10$^{9}$) & ($g/cm^{2}$)     &($g/cm^{2}$) & (EeV)&  (EeV)& (\%) \\                                                                    
\hline                                                                                                                            
 Event1  & 6.4 & 0.93     & 1.06   & 701 & 706 & 1.370 &1.562 & 14\\
 Event2    & 8   & 6.57     & 6.88   & 759 & 767 & 9.853&10.40 & 6\\
 Event3   & 11  & 2.12     & 2.19   & 637 &642 &2.950 &3.100 &5\\
 Event4  & 23  & 12.85     & 13.10   & 752 & 753 &19.20 &19.57 & 2  \\
                                                               
\hline                                                                                                                            
\hline                                                                                                                            
\end{tabular}                                                                                                                     
\end{table}                                                                                                                       

\begin{figure}[ht]
\begin{center}
\includegraphics[height=11cm,width=12cm,angle=0]{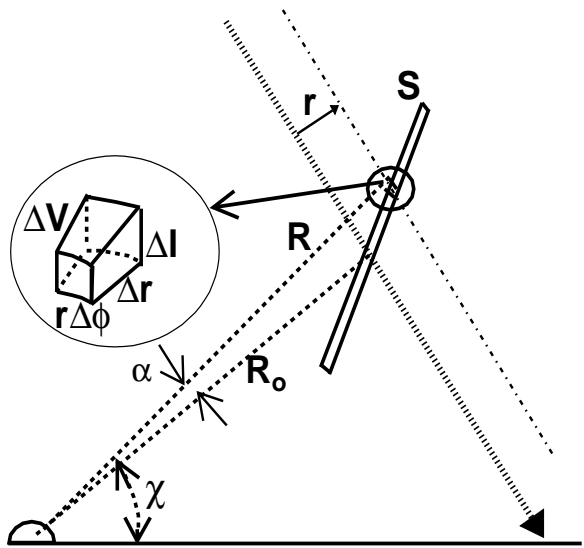}
\end{center}
\caption{ {\it Geometry  of an EAS as seen by the fluorescence detector.
Photons which arrive simultaneously at the FD originate from the  surface S. See text for more details.
}}
\label{fig1}
\end{figure}
\begin{figure}[ht]
\begin{center}
\includegraphics[height=13.5cm,width=8.5cm,angle=-90]{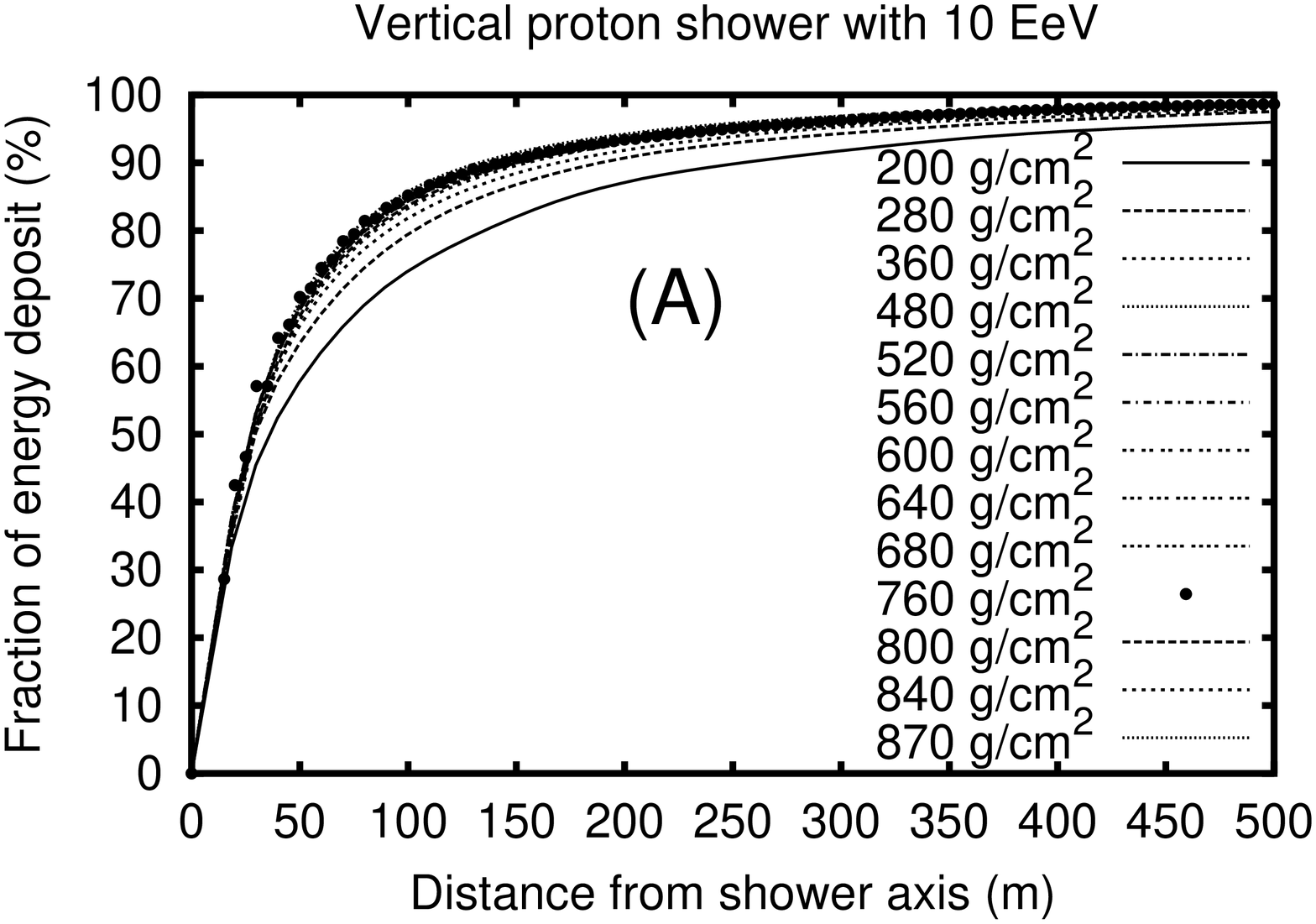}

\includegraphics[height=13.5cm,width=8.5cm,angle=-90]{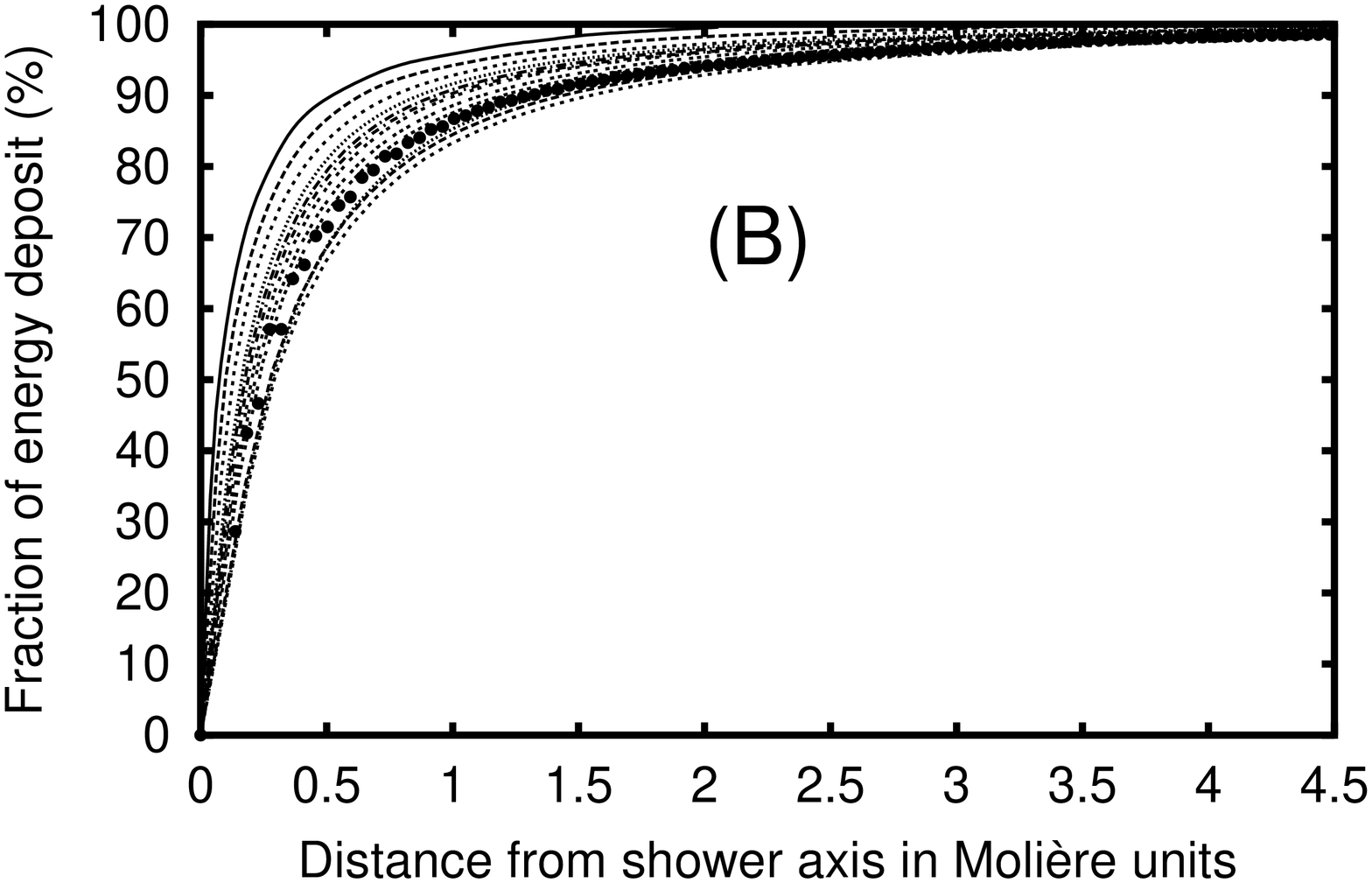}

\end{center}
\vspace{0.5 cm}
\caption{{\it (A) Integral of energy deposit density  $F(r)$ versus distance from shower axis for
  proton  shower; (B) Integral of energy deposit density $F(r^{*})$ versus distance 
  from shower axis measured in the Moli\`ere units.
  Different lines described in the top panel  correspond to profiles obtained for different vertical atmospheric
  depths. The  profile corresponding to shower maximum is marked by dots. 
}}
\label{fig339int}
\end{figure}

\begin{figure}[t]
\begin{center}
\includegraphics[height=13.5cm,width=8.5cm,angle=-90]{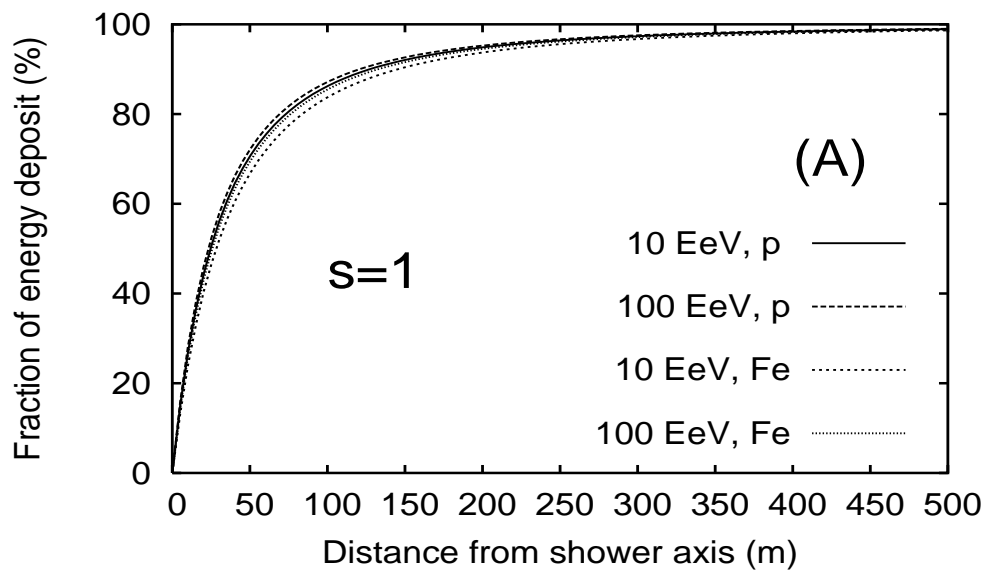}

\includegraphics[height=13.5cm,width=8.5cm,angle=-90]{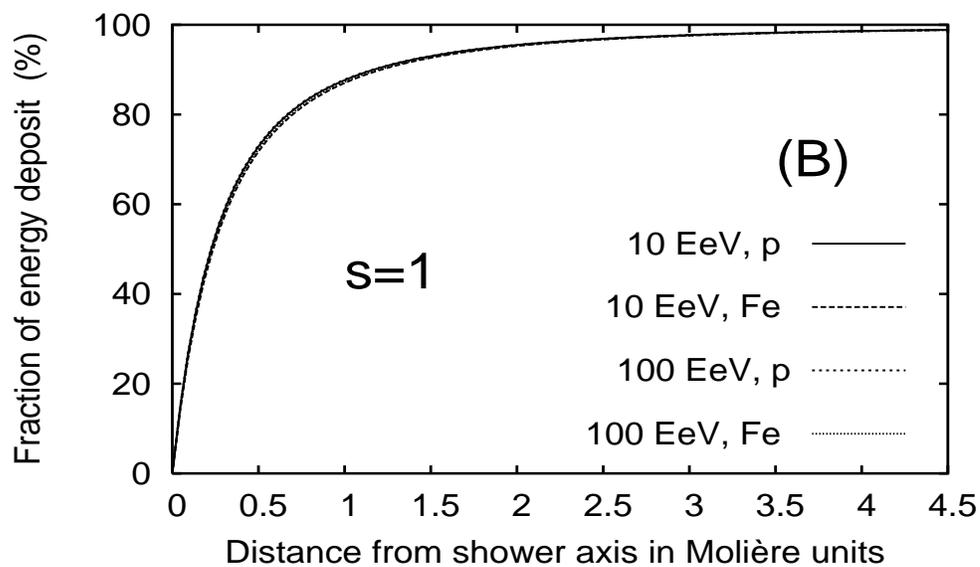}
\end{center}
\caption{ {\it (A) Integral of   energy deposit density  versus distance
 from shower axis; 
(B) The integral profiles versus distance   measured in Moli\`ere units; 
 The profiles are shown for vertical showers (at $s$=1) with  
 different primary particle type and  energy.   
}}
\label{fig-en}
\end{figure}

\begin{figure}[t]
\begin{center}
\includegraphics[height=13.5cm,width=8.5cm,angle=-90]{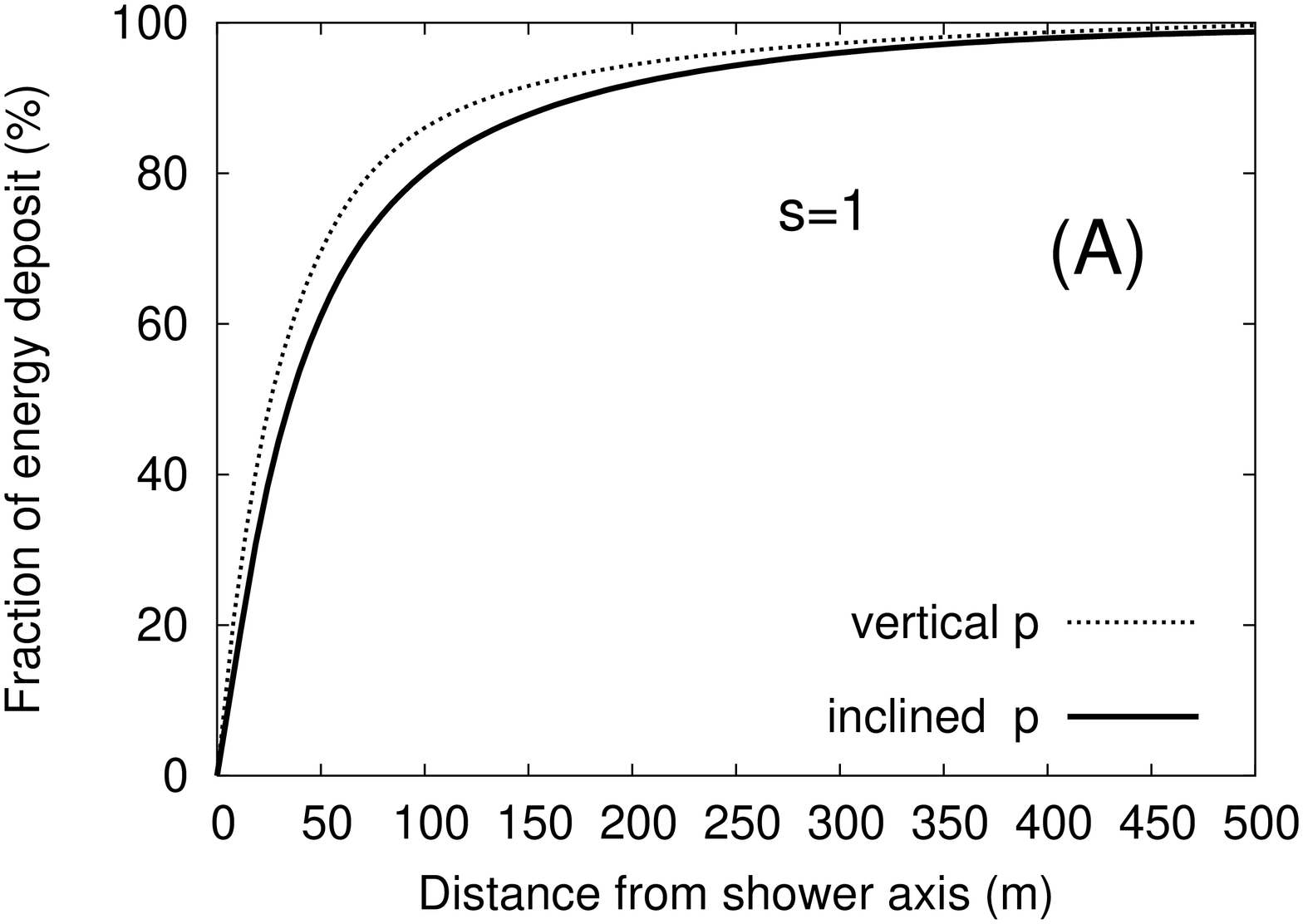}

\includegraphics[height=13.5cm,width=8.5cm,angle=-90]{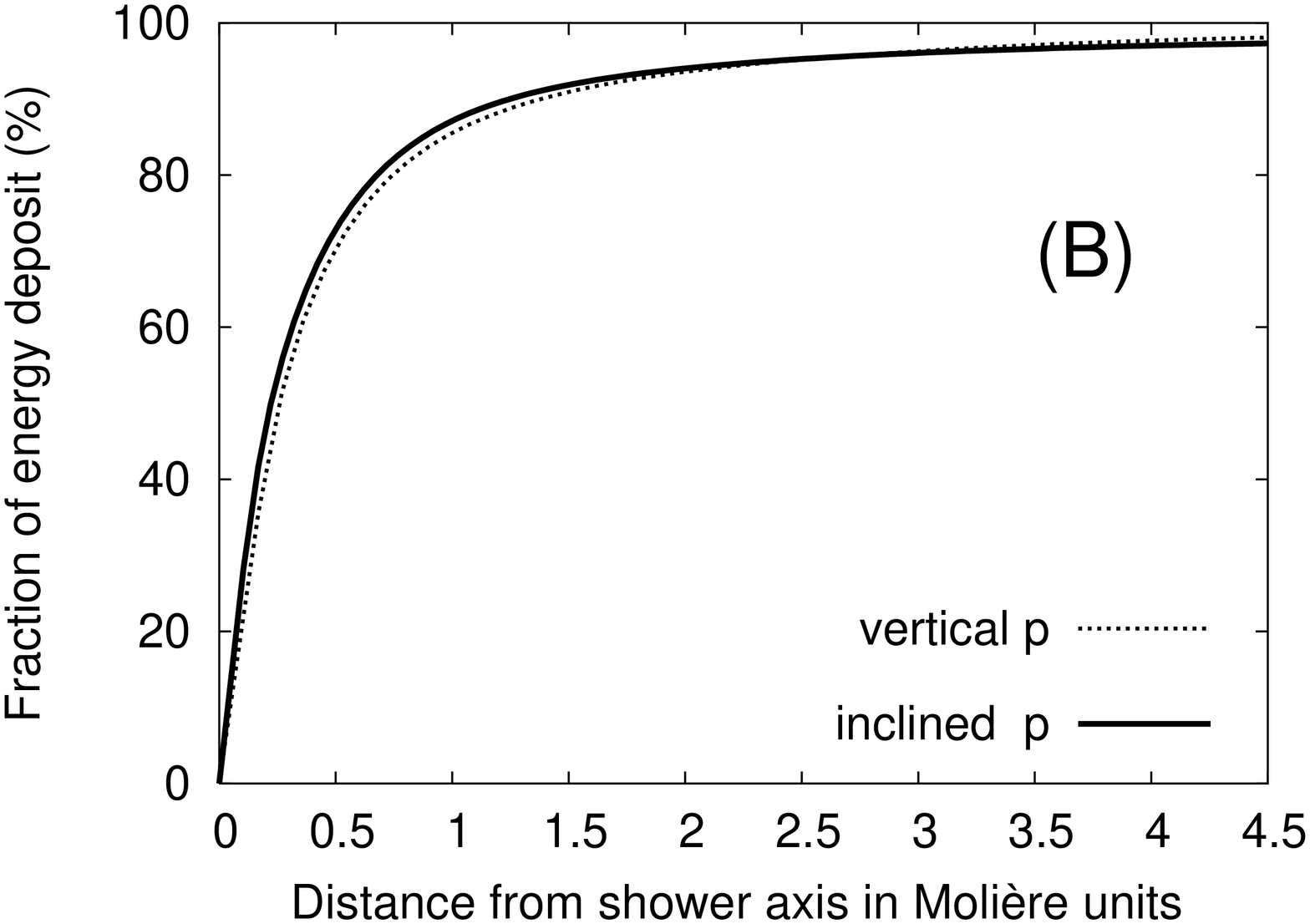}

\end{center}
\caption{ {\it (A)  Integral of   energy deposit density versus distance from shower axis for vertical  
and inclined ($\theta=45^{\circ}$) proton showers; (B) The integral profile 
measured in Moli\`ere units. The profiles are shown for   10 EeV showers at $s=1$.
}}
\label{fig-inc}
\end{figure}

\begin{figure}[t]
\begin{center}
\includegraphics[height=6.5cm,width=6.0cm,angle=-90]{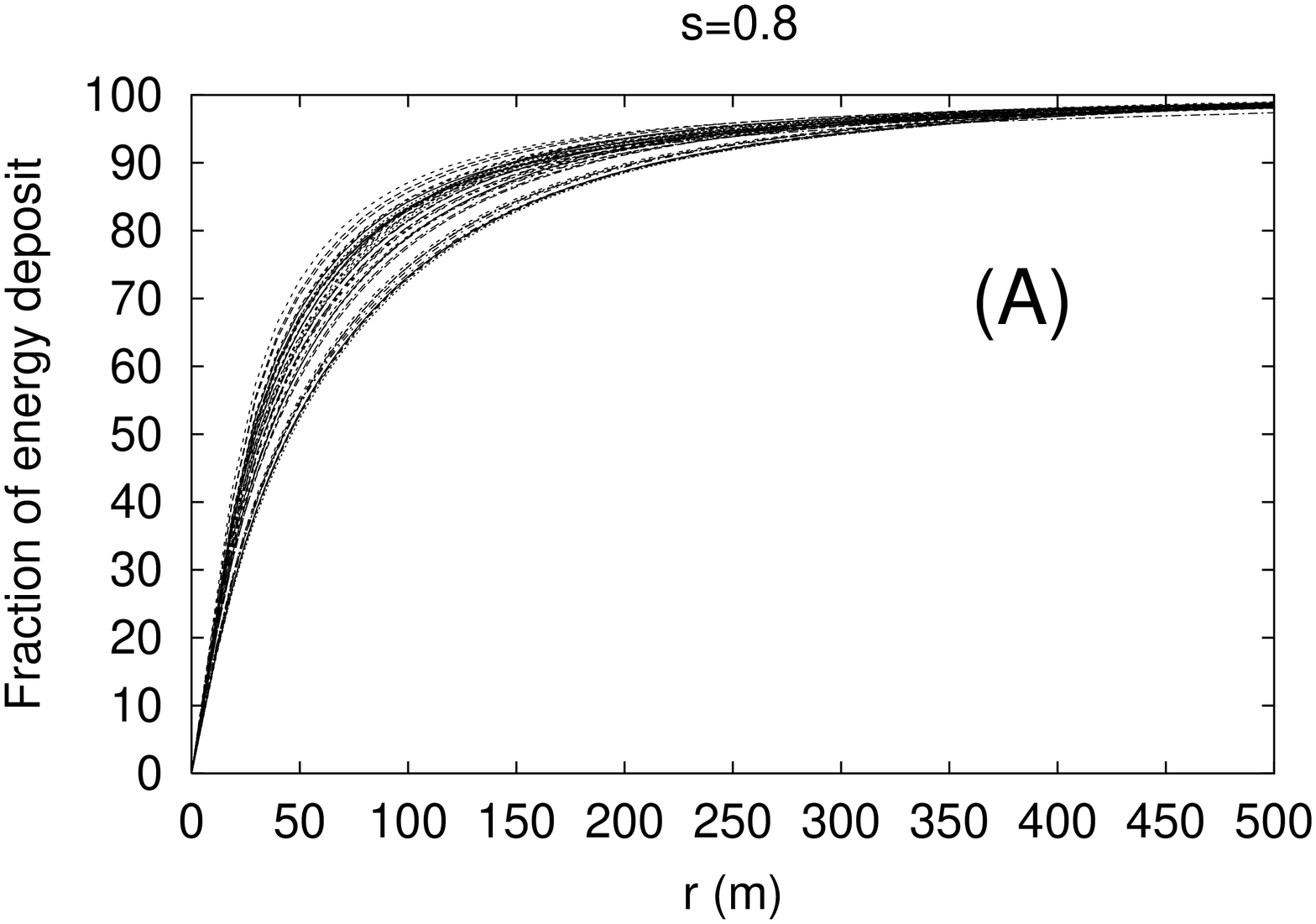}
\includegraphics[height=6.5cm,width=6.0cm,angle=-90]{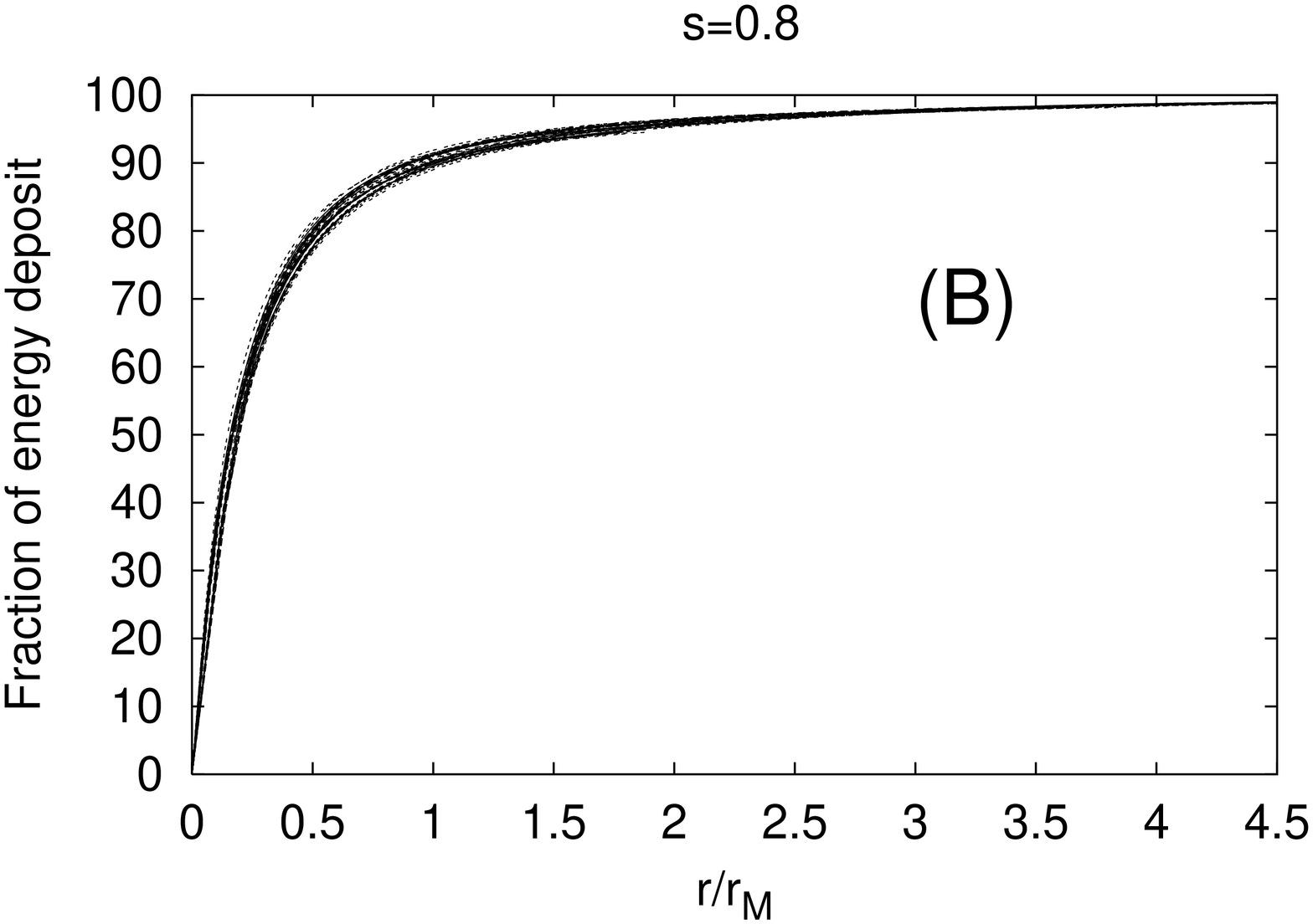}
\includegraphics[height=6.5cm,width=6.0cm,angle=-90]{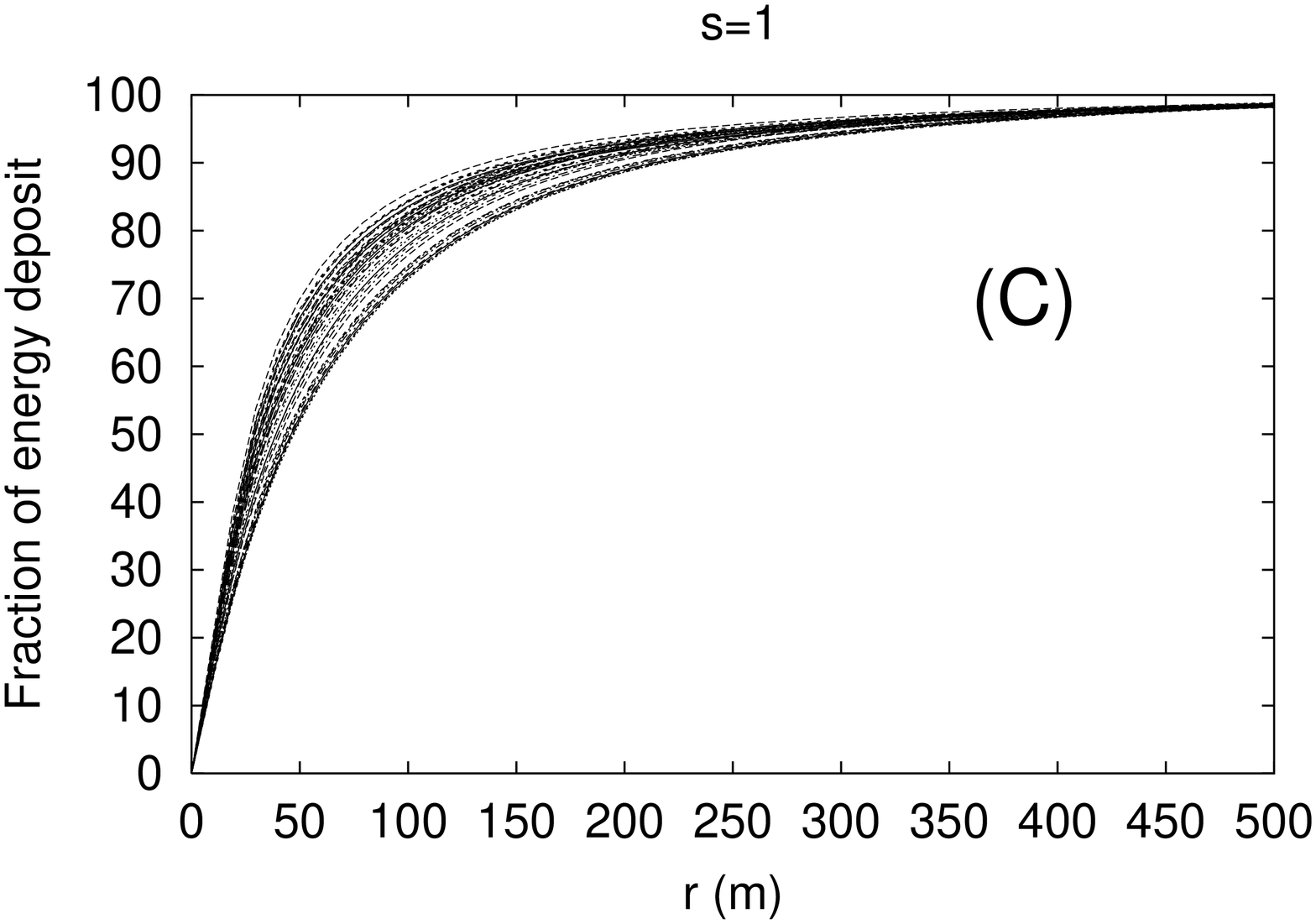}
\includegraphics[height=6.5cm,width=6.0cm,angle=-90]{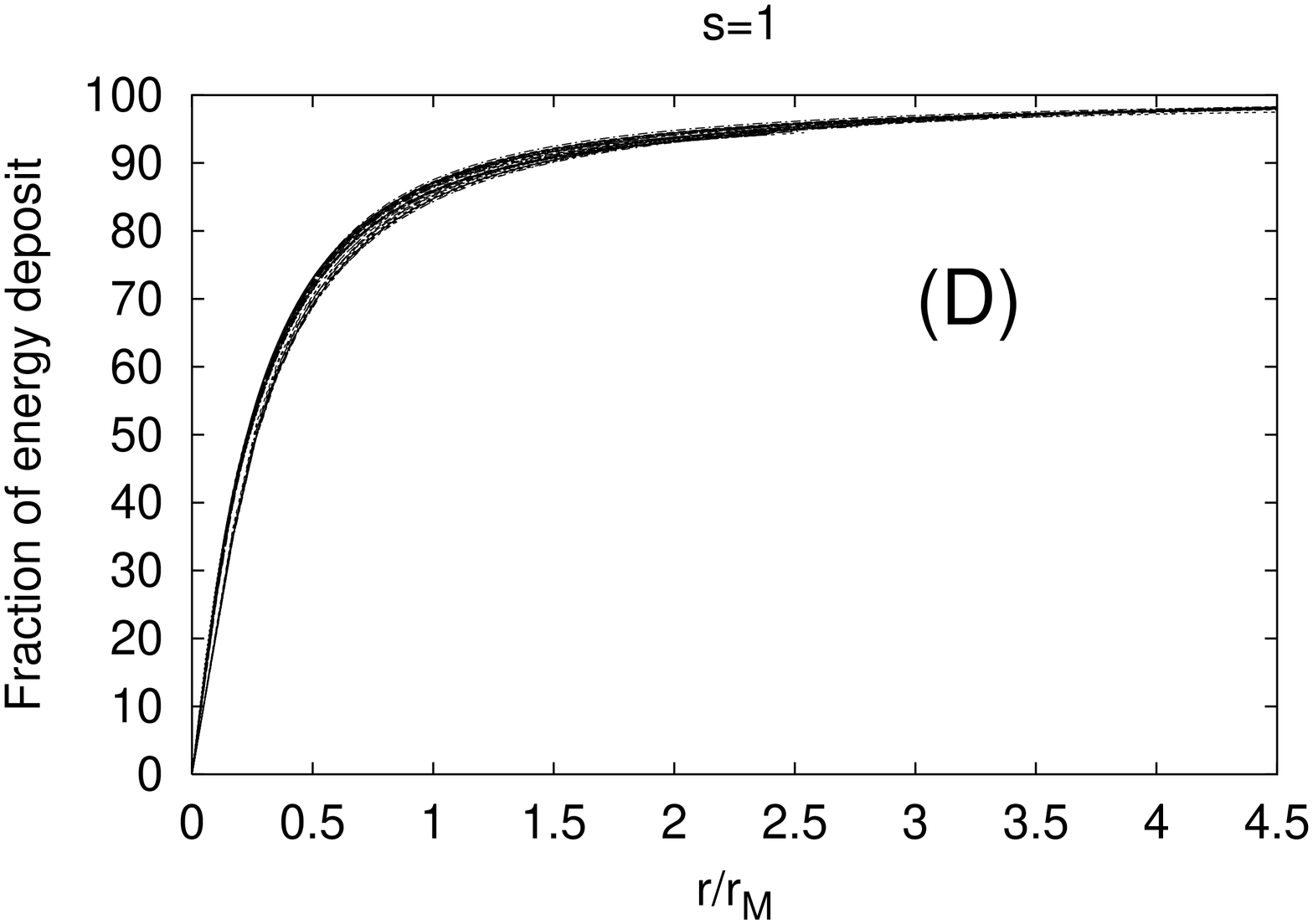}
\includegraphics[height=6.5cm,width=6.0cm,angle=-90]{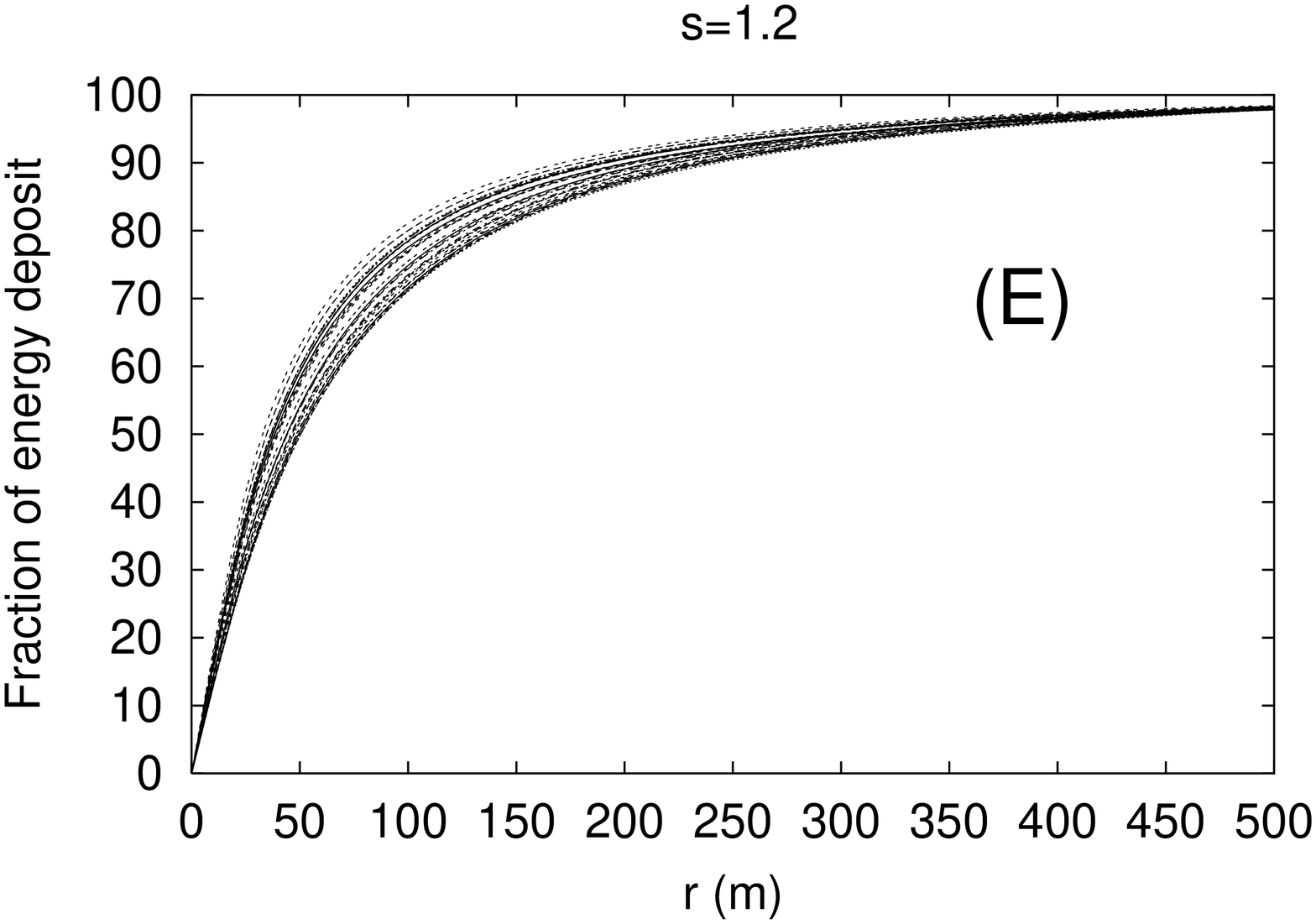}
\includegraphics[height=6.5cm,width=6.0cm,angle=-90]{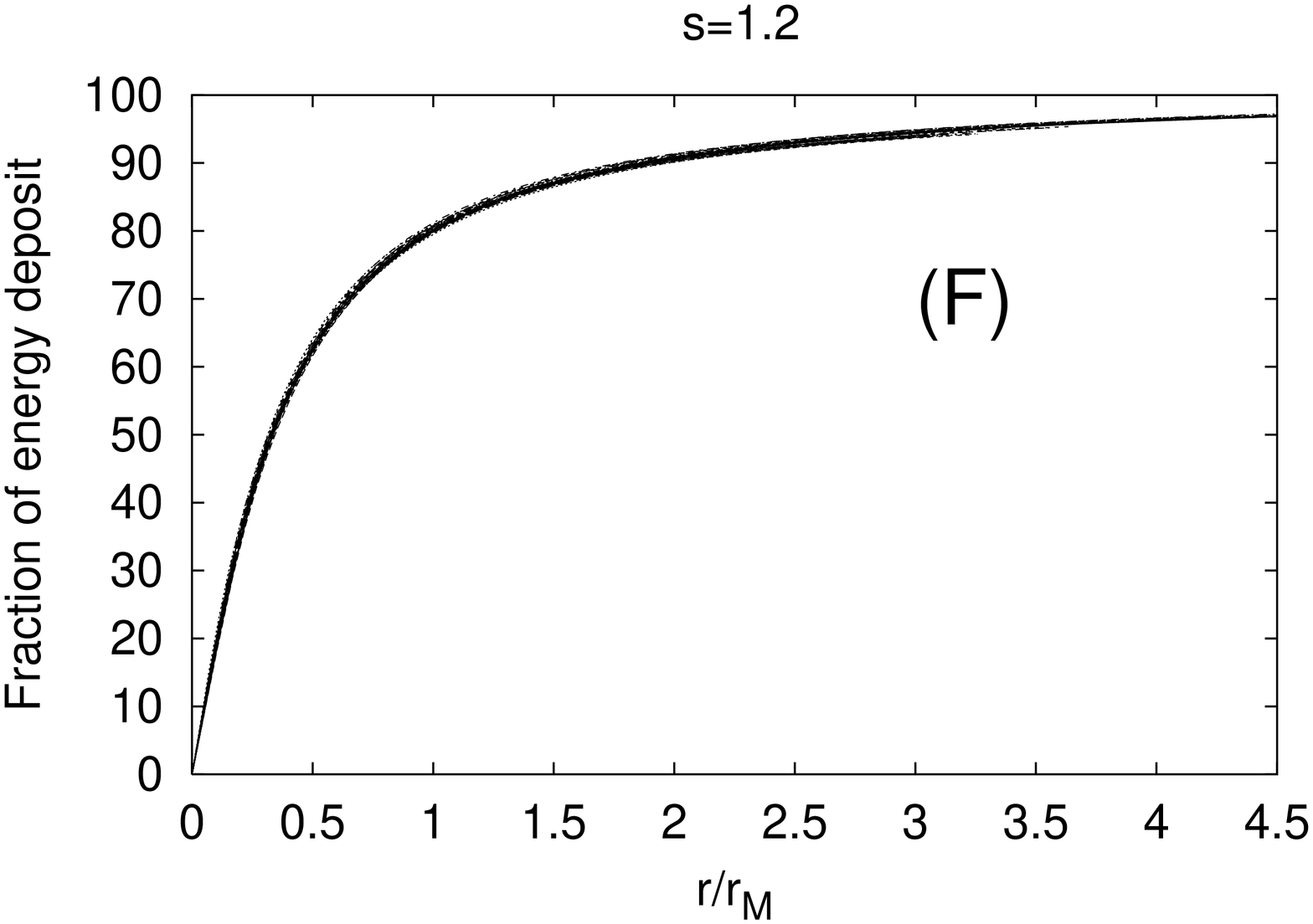}

\end{center}
\vspace{0.5 cm}
\caption{{\it (A), (C), (E) Integral of energy deposit density versus distance from shower axis;
(B), (D), (F) Integral of energy deposit density versus distance measured in Moli\`ere units.
 Individual   profiles  are obtained for showers with  different age, zenith angle
and primary particles.
}}
\label{fig9int}
\end{figure}

\begin{figure}[t]
\begin{center}
\includegraphics[height=13.5cm,width=8.5cm,angle=-90]{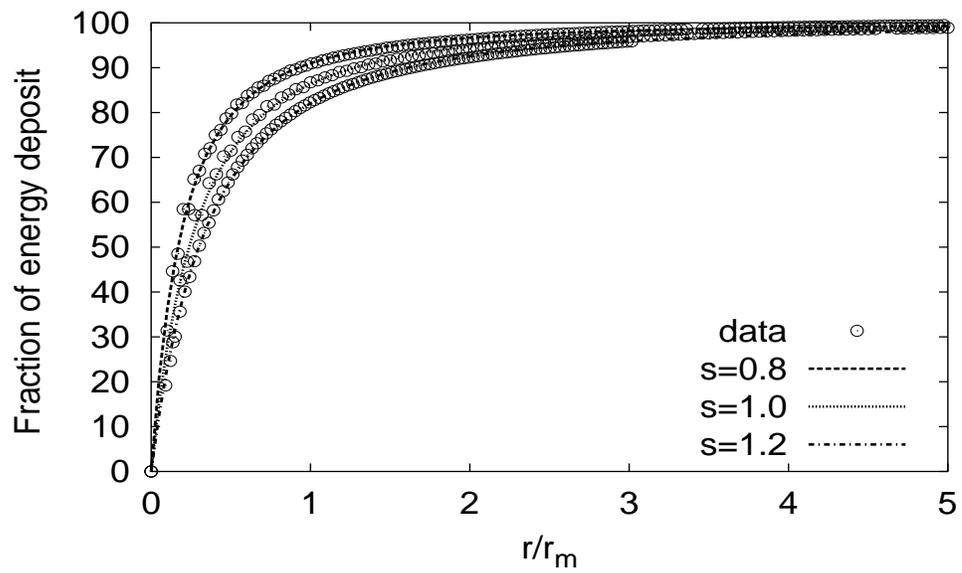}

\end{center}
\caption{ {\it The fit of function Eq.~(\ref{eq-fit}) to fraction of energy deposit density 
from Figure~\ref{fig9int}B, D, F
for different values of the age parameter $s$.
}}
\label{fit-age}
\end{figure}

\begin{figure}[t]                                                                                         
\begin{center}                                                                                            
\includegraphics[height=13.5cm,width=8.5cm,angle=-90]{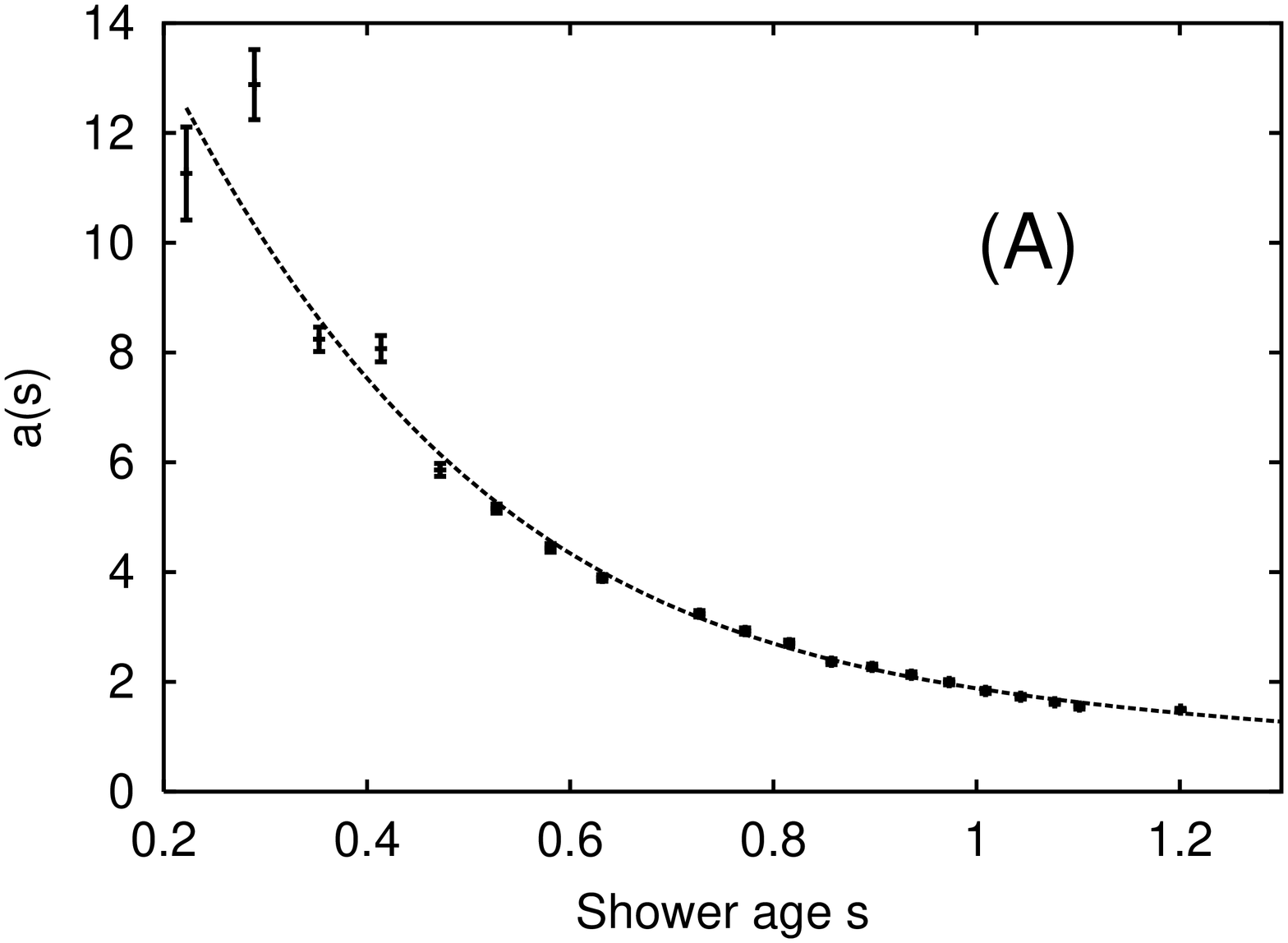}                                                 
                                                                                                          
\includegraphics[height=13.5cm,width=8.5cm,angle=-90]{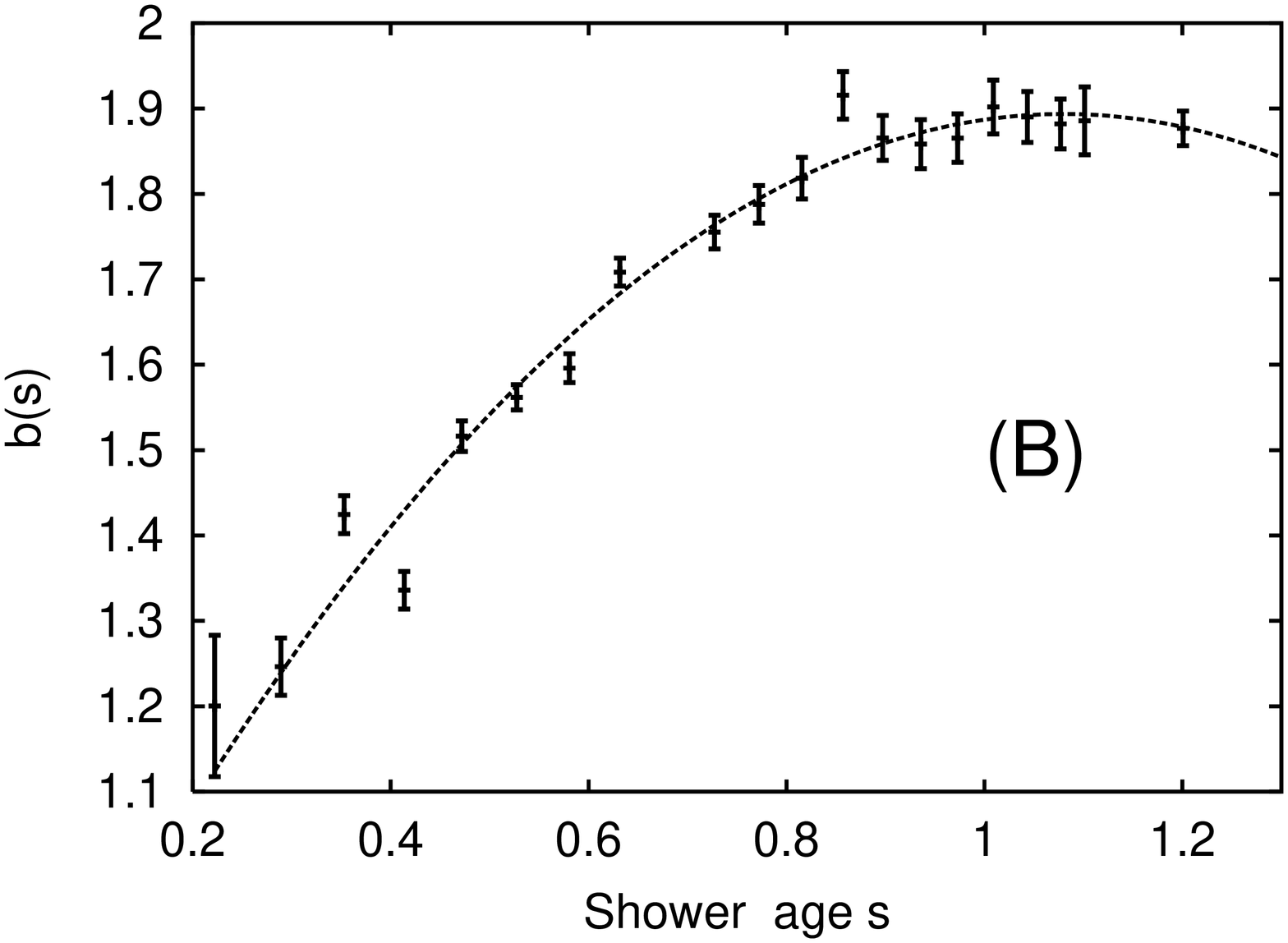}

\end{center}                                                                                              
\caption{ {\it Values of parameters $a(s)$   and $b(s)$ of Eqs.~(8) and (9) obtained based on integral                         
of CORSIKA  energy deposit density for vertical  showers  at energy 10 EeV.  
}}                                                                                                        
\label{fig-par}                                                                                           
\end{figure}          

 \begin{figure}[t]
\begin{center}

\includegraphics[height=13.0cm,width=8.5cm,angle=-90]{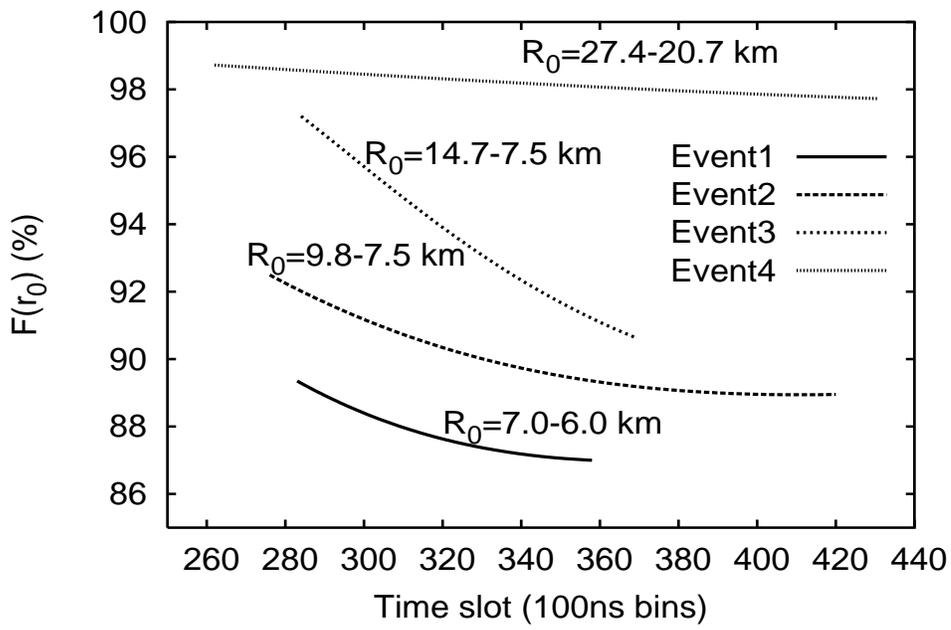}
\end{center}
\vspace{0.5 cm}
\caption{{\it  Fraction of light collected within the angle $\chi_{S/N}^{max}$  versus time for some simulated 
  events.
See text for more details.
}}
\label{fig-frac}
\end{figure}

\begin{figure}[t]
\begin{center}

\includegraphics[height=13.5cm,width=8.5cm,angle=-90]{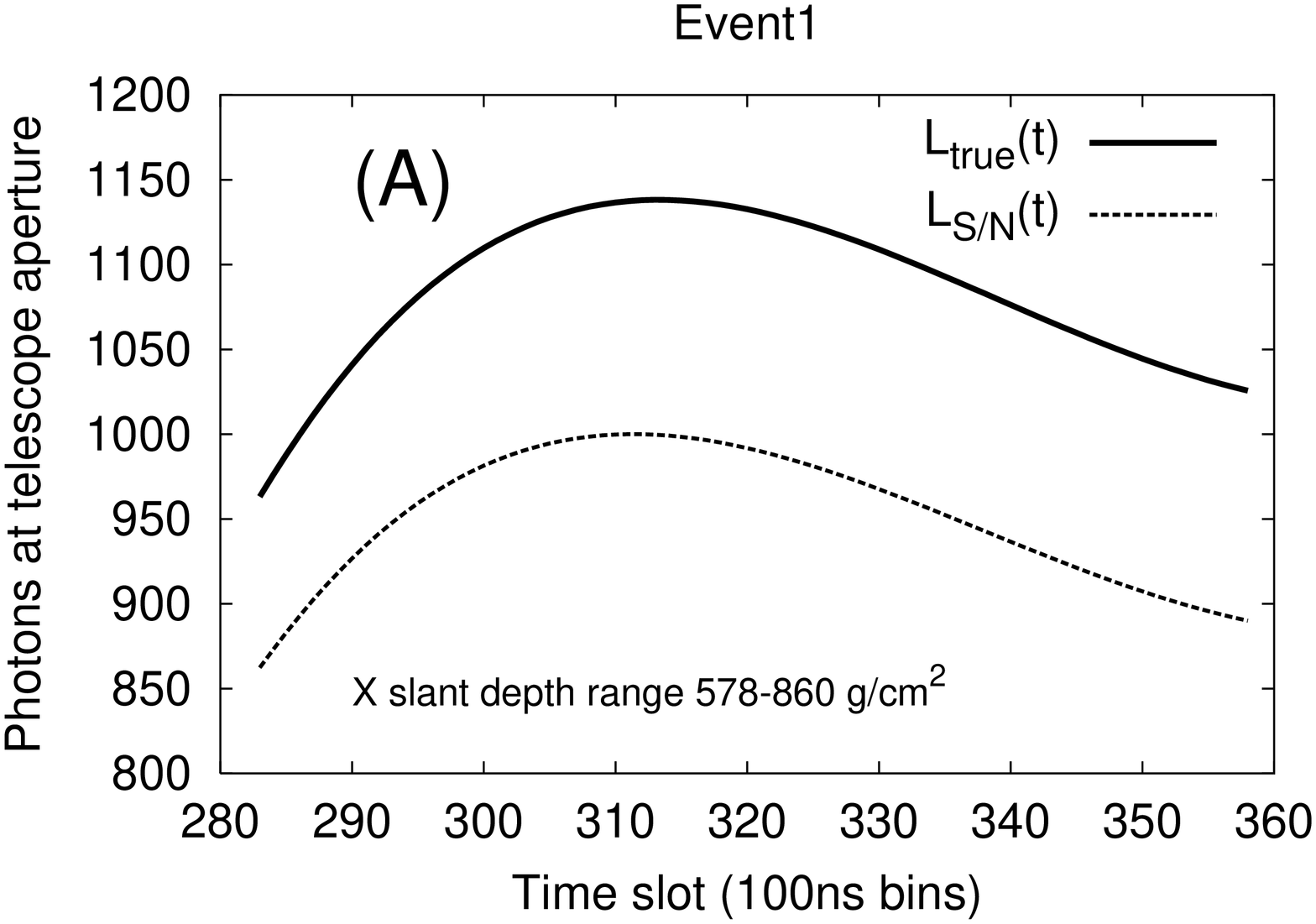}

\includegraphics[height=13.5cm,width=8.5cm,angle=-90]{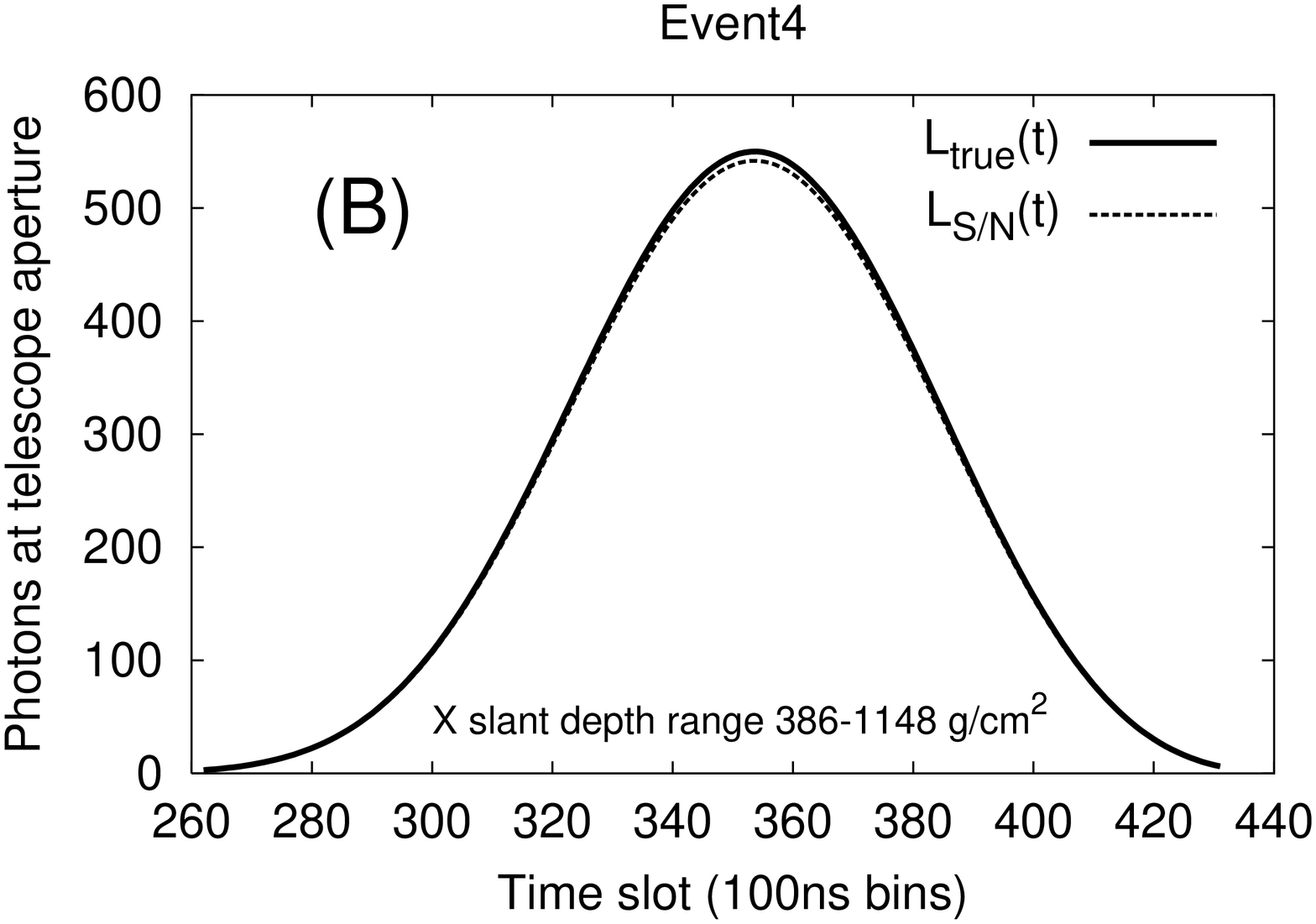}

\end{center}
\vspace{0.5 cm}

\caption{{\it(A) Comparison of light profiles versus time slot number for  Event1.
The  $L_{S/N}(t)$ light profile (dashed line) is  obtained based on  
the signal-to-noise algorithm implemented in FloresEye.
The $L_{true}(t)$  light profile (solid line) is  obtained using    the 
shape of $F(r)$ function  derived from Eq.~(\ref{eq-fit}).   
(B)  Comparison of  light profile versus time slot number for Event4.
}}
\label{fig-prof2}
\end{figure}

\end{document}